\documentclass[aps,prl,twocolumn,amsmath]{revtex4-2}
\usepackage{bm}
\usepackage{graphicx}

\usepackage{pdfpages}
\usepackage{etoolbox} 

\makeatletter
\patchcmd{\@outputpage@head}{\@ifx{\LS@rot\@undefined}{}{\LS@rot}}{}{}{}
\makeatother

\begin{document}

\title{Superconducting insulators and localization of Cooper pairs}
\author{Konstantin Yu. Arutyunov$^{1,2}$, Janne S. Lehtinen$^3$, Alexey Radkevich$^4$, Andrew G. Semenov$^{1,4}$
and Andrei D. Zaikin$^{5,4}$}
\affiliation{$^1$National Research University Higher School of Economics, 101000 Moscow, Russia\\
$^2$P.L. Kapitza Institute for Physical Problems RAS, 119334, Moscow, Russia\\
$^3$VTT Technical Research Centre of Finland Ltd., Centre for Metrology MIKES, P.O. Box 1000, FI-02044 VTT, Espoo, Finland\\
$^4$I.E.Tamm Department of Theoretical Physics, P.N.Lebedev Physical Institute, 119991 Moscow, Russia\\
$^5$Institute for Quantum Materials and Technologies, Karlsruhe Institute of Technology (KIT), 76021 Karlsruhe, Germany}

\date{\today}
\begin{abstract}
{\bf Abstract}

Rapid miniaturization of electronic devices and circuits demands profound understanding of fluctuation phenomena at the nanoscale. Superconducting nanowires -- serving as important building blocks for such devices -- may seriously suffer from fluctuations which tend to destroy long-range order and suppress superconductivity. In particular, quantum phase slips (QPS) proliferating at low temperatures may
turn a quasi-one-dimensional superconductor into a resistor or an insulator. Here, we introduce a physical concept of QPS-controlled  localization of Cooper pairs that may occur even in uniform nanowires without any dielectric barriers being a fundamental manifestation of the flux-charge duality in superconductors. We demonstrate -- both experimentally and theoretically -- that deep in the "insulating" state such nanowires actually exhibit non-trivial superposition of superconductivity and weak Coulomb blockade of Cooper pairs generated by quantum tunneling of magnetic fluxons across the wire.
\end{abstract}
%\pacs{}
\maketitle

{\bf INTRODUCTION}

Superconducting nanowires represent an important example of a system where low temperature physics is dominated by both thermal and quantum fluctuations \cite{book,AGZ,LV,Bezrbook,SZ20}, thus making their properties entirely different from those of bulk superconductors well described by the standard Bardeen-Cooper-Schriffer (BCS) mean field theory \cite{Tinkham}. 

A large part of fluctuation phenomena in such nanowires are attributed to the so-called phase slips \cite{book,AGZ} which correspond to temporal local suppression of the superconducting order parameter $\Delta \exp (i \varphi )$ accompanied by the phase slippage process. At temperatures $T$ close enough to the BCS critical temperature $T_C$ such phase slips are induced by thermal fluctuations \cite{Little,MH,GZTAPS} whereas at lower temperatures $T \ll T_C$ quantum fluctuations of the order parameter take over and generate quantum phase slips (QPS) \cite{ZGOZ,GZQPS}.

As the phase $\varphi$ changes in time by $2\pi$ during a QPS event, each such event causes a voltage pulse $V=\dot{\varphi}/2e$ inside the wire. As a result, a current biased superconducting nanowire acquires a non-vanishing electric resistance down to lowest $T$ \cite{ZGOZ,GZQPS}. This effect received its convincing experimental confirmation \cite{BT,Lau,Zgi08,Leh,LA,liege,Aetal}. The same effect is also responsible for voltage fluctuations in superconducting nanowires \cite{SZ16,SZ19}. Quantum phase slips also cause suppression of persistent currents in uniform superconducting nanorings \cite{KoAr,SZ13}.
\begin{figure}
\begin{center}
\includegraphics[width=80mm]{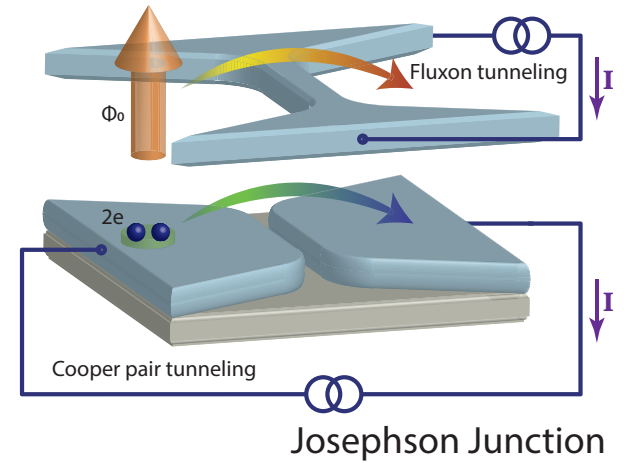}
\end{center}
\caption{{\bf Flux-charge duality.} Quantum tunneling of a magnetic fluxon $\Phi_0$ across a superconducting nanowire and a dual tunneling process of a Cooper pair with charge $2e$ across a Josephson junction between two superconductors.}
\end{figure}

A fundamentally important property of superconducting nanowires is the so-called flux-charge duality. This feature was extensively discussed for ultrasmall Josephson junctions \cite{ZP87,averin,Z90,SZ90} implying that under the duality transformation $2e \leftrightarrow \Phi_0$ quantum dynamics of Cooper pairs (with charge $2e$) should be identical to that of magnetic flux quanta $\Phi_0=hc/2e$. All the same arguments remain applicable for shorter superconducting nanowires \cite{MN} which properties are dual to those of small Josephson junctions (Fig. 1). The duality considerations can further be extended to longer nanowires \cite{SZ20,SZ13}. 

Manifestations of flux-charge duality in superconducting nanowires were observed in a variety of experiments thereby opening new horizons for applications of such structures in modern nanoelectronics, information technology and metrology. These observations include, e.g., coherent tunneling of magnetic flux quanta through superconducting nanowires \cite{AstNature,AstPRB} and the so-called Bloch steps \cite{Kostya2}. Operations of duality-based single-charge transistor \cite{Zorin,KA} and charge quantum interference device \cite{Zhenya} were demonstrated. Superconducting nanowires were also proposed to serve as central elements for QPS flux qubits \cite{Hans} as well as for creating a QPS-based standard of electric current \cite{Wang}. 

Quantum fluctuations in superconducting nanowires are controlled by two different parameters
\begin{equation}
g_\xi = R_q/R_\xi \;\;\;\; {\rm and} \;\;\; g_Z=R_q/Z.
\label{1}
\end{equation}
Here $R_q =h/e^2 \simeq 25.8$ K$\Omega$ is the quantum resistance unit, $R_\xi$ is the normal state resistance of the wire segment of length equal to the superconducting coherence length $\xi$ and $Z=\sqrt{\mathcal{L}/C}$ is the wire impedance determined by the kinetic wire inductance (times length) $\mathcal{L}$ and the geometric wire capacitance (per length) $C$. 

The dimensionless conductance $g_\xi$ accounts for the fluctuation correction to the BCS order parameter \cite{GZTAPS} $\Delta\to \Delta -\delta \Delta$ (with $\delta \Delta  \sim \Delta /g_\xi$) and determines the QPS amplitude (per unit wire length) \cite{GZQPS} $\gamma_{\rm qps} =b(g_\xi\Delta/\xi)\exp (-a g_\xi)$ (with $a \sim 1$ and $b \sim 1$). The dimensionless admittance $g_Z$, in turn, accounts for hydrodynamic (long wavelength) fluctuations of the superconducting phase intimately related to sound-like plasma modes \cite{MS} propagating along the wire with the velocity $v=1/\sqrt{\mathcal{L}C}$. Different quantum phase slips interact by exchanging such plasmons and, hence, the parameter $g_Z$ also controls the strength of inter-QPS interactions. By reducing the wire diameter $\sqrt{s} \propto g_Z$  one eventually arrives at the "superconductor-insulator" quantum phase transition \cite{ZGOZ} that occurs at $g_Z=16$ and $T \to 0$. 

In this work, we experimentally and theoretically investigate both global and local ground state properties of superconducting nanowires in the "insulating" regime $g_Z<16$. We demonstrate that quantum fluctuations of magnetic flux in long nanowires yield effective localization of Cooper pairs at a fundamental length scale $L_c$ that essentially depends on both parameters (\ref{1}).  We also show that nominally uniform nanowires exhibit a non-trivial mixture of superconducting-like features at shorter length scales and resistive long-scale behavior which should actually tend to insulating at $T \to 0$. This state of matter can thus be named as a superconducting insulator.

\begin{figure}
\begin{center}
\includegraphics[width=80mm]{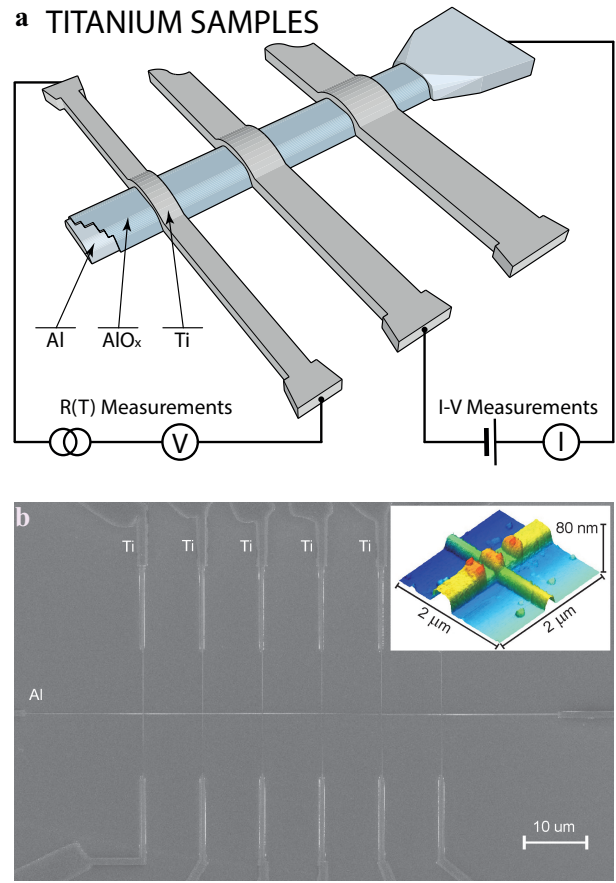}
\end{center}
\caption{{\bf Schematics of experiment and sample layout.} (a) Long and thin titanium nanowires having the form of narrow strips overlap a relatively wide aluminum electrode through a tunnel barrier (aluminum oxide). The structure enables one to carry out both pseudo-four-terminal measurements of the total resistance for all nanowires and local measurements of the current-voltage characteristics for all $Al-AlOx-Ti$ tunnel junctions. (b) Scanning electron microscope image of our typical structure. Inset: Zoom of the junction region taken with atomic force microscope. Fake color corresponds to variation of the sample height from 0 (substrate, dark blue) to 80 nm (overlapping titanium, orange).}
\end{figure}

{\bf RESULTS}

In order to accomplish our goal we fabricated long and thin titanium nanowires having the form of narrow strips overlapping a relatively wide aluminum electrode through a tunnel barrier (aluminum oxide), as it is shown in Fig. 2. The normal state resistance of these wires $R_N$ measured above the BCS critical temperature $T_C \approx 400$ mK was found in the range $R_N \sim 25 \div 70$ k$\Omega$. The length $L \simeq$ 20 $\mu$m and thickness $d\simeq 35$ nm remain the same for all $Ti$ samples, whereas their width $w$ varies in the range $30 \div 60$ nm within which quantum phase slips usually proliferate in $Ti$ nanowires \cite{Leh,KK}. The zero temperature superconducting coherence length in our $Ti$ samples are estimated to be $\xi \sim 140 \div150$ nm and, hence, the quasi-one-dimensional limit condition $d,w \ll \xi \ll L$ holds for all samples. With these parameters, one obtains the dimensionless admittance $g_Z \approx 1 \div 3$, i.e. the desired condition $g_Z <16$ is well satisfied in all our nanowires. The dimensions of the aluminum strip are large enough enabling one to ignore fluctuation effects. 

{\bf Nanowire resistance}

The results of our measurements of a total resistance $R(T)$ for five different nanowires are displayed in Fig. 3a. With the values $g_Z \ll 16$,  in the low temperature limit all these samples should remain deep in the insulating regime. We observe, however, that two thicker samples with nominal widths $w \approx 62$ nm (sample $Ti1$) and $w \approx 46$ nm (sample $Ti2$), demonstrate a pronounced resistive behavior with $R(T)\approx R_N$ only at temperatures not far below the bulk titanium critical temperature  $T_C \approx 400$ mK followed by a rather sharp resistance drop by $\sim 2$ orders of magnitude at temperatures $T \sim 300$ mK (sample $Ti1$) and $T \lesssim 200$ mK (sample $Ti2$).   The remaining samples $Ti3$, $Ti4$ and $Ti5$ with nominal widths just slightly below that for $Ti2$ (respectively $w \approx 41$ nm, 40 nm and 30 nm) show no sign of superconductivity down to the lowest $T$ and only very weak dependence $R(T)$, in particular for the thinnest samples $Ti4$ and $Ti5$. 

\begin{figure}
\includegraphics[width=80mm]{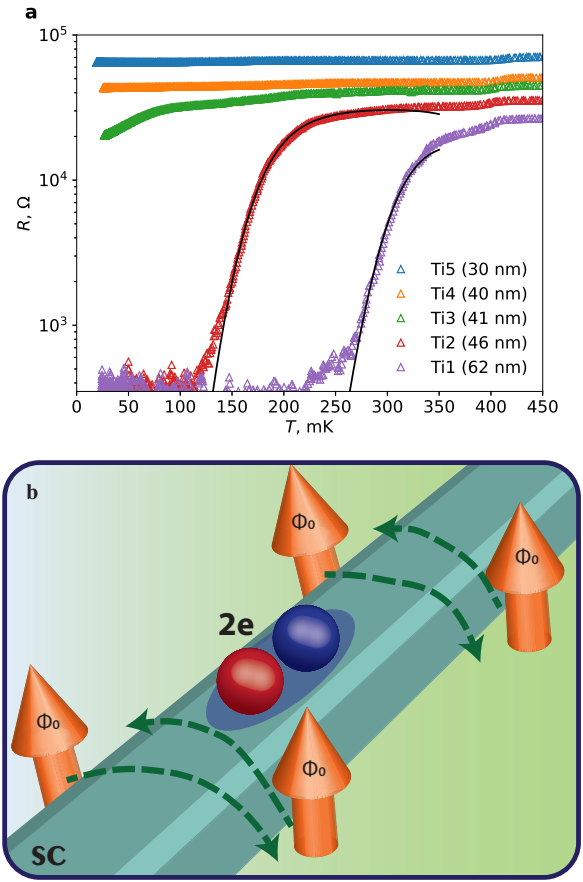}
\caption{{\bf Resistance data and localization of Cooper pairs.} (a) Resistance versus temperature $R(T)$ measured for five $Ti$ nanowires of length $L=20$ $\mu$m, thickness $d\simeq 35$ nm and nominal width values $w$ indicated in brackets for each of the samples. Solid lines represent fits of the data for samples $Ti1$ and $Ti2$ to the theory \cite{GZTAPS} within its validity range. Resistance saturation observed in these two samples at low $T$ is due to finite voltage sensitivity of about few nV corresponding to residual resistance $\sim 100$ $\Omega$ measured using ac bias current $\sim 10$ pA rms. Error bars are smaller than data points. (b) Localization of Cooper pairs (with charge $2e$) generated by quantum tunneling of magnetic fluxons $\Phi_0$ across the nanowire. This phenomenon explains the low temperature behavior of $R(T)$ observed in samples $Ti3$, $Ti4$ and $Ti5$.}
\end{figure}

At temperatures not far below $T_C$ the system behavior should be dominated by thermally activated phase slips which contribution to the wire resistance $R_{\rm taps}(T)$  \cite{GZTAPS} indeed provides very accurate fits for the resistance of two of the above samples  (see Fig. 3a) and allows to extract effective values $g_\xi \simeq 37.4$ and $g_\xi \simeq 9.0$ respectively for samples $Ti1$ and $Ti2$ (see Supplementary Note 1 for more details). These values are smaller than the nominal ones, most likely indicating certain non-uniformity of our nanowires.

{\bf Localization of Cooper pairs}

In order to understand drastic difference in the low temperature behavior of our samples with various cross sections it is necessary to account for the effect of quantum phase slips. The dual Hamiltonian for superconducting nanowires in the presence of QPS reads \cite{SZ20,SZ13}
\begin{equation}
\hat H=\int_0^Ldx \left[\frac{\hat\Phi^2}{2{\mathcal L}}+\frac{(\partial_x \hat Q)^2}{2C}-\gamma_{\rm qps}\cos \left(\frac{\pi \hat Q}{e}\right)\right],
\label{H}
\end{equation}
where $\hat\Phi$ and $\hat Q$ are canonically conjugate flux and charge operators obeying the commutation relation $[\hat\Phi (x),\hat Q (x')]=- i \hbar\delta (x-x')$. Employing this Hamiltonian one can demonstrate  \cite{SZ20,SZ13} that in the "insulating" phase, i.e. for $g_Z<16$, the wire ground state properties are controlled by a non-perturbative correlation length $L_c \propto \gamma_{\rm qps}^{-\alpha}$ with $1/\alpha =2-g_Z/8$ or, equivalently,
\begin{equation}
L_c \sim \xi \exp\left(\frac{ag_\xi-\ln b}{2-g_Z/8}\right).
\label{LQPS}
\end{equation}
Physically the appearance of this QPS-induced fundamental length scale can be viewed as a result of spontaneous tunneling of magnetic flux quanta $\Phi_0$ back and forth across the wire, as it is illustrated in Fig. 3b. These quantum fluctuations of magnetic flux wipe out phase coherence at distances $\sim L_c$ and yield effective localization of Cooper pairs at such length scales.   Accordingly, samples with $L \lesssim L_c$ may still exhibit superconducting properties also in the presence of QPS, whereas in the limit $L \gg L_c$ the supercurrent gets disrupted by quantum fluctuations and such nanowires remain non-superconducting even at $T\to 0$.

This is exactly what the data in Fig. 3a demonstrate. Indeed, the value $L_c$ (\ref{LQPS}) for the sample $Ti1$ with $g_\xi \approx 37$ obviously exceeds $L$ by several orders of magnitude and, hence, this sample should remain superconducting at low enough $T$. In order to estimate the length scale  (\ref{LQPS}) for sample $Ti2$ with $\xi \sim 140$ nm, $g_\xi \simeq 9.0$ and $g_Z \simeq 2.5$ it is desirable to explicitly determine the prefactors $a$ and $b$. The data analysis for this sample yields a lower bound for the combination $a g_\xi -\ln b \gtrsim 7.5$, see Supplementary Note 2 for details. With this in mind Eq. (\ref{LQPS}) allows to estimate $L_c \gtrsim 12$ $\mu$m, i.e. in this case $L_c \sim L$ and the sample $Ti2$ should also remain superconducting at low $T$ in accordance with our observations. By contrast, three thinner nanowires $Ti3$, $Ti4$ and $Ti5$ with lower effective values $g_\xi$ and $L_c$ significantly smaller than $L$ exhibit a non-superconducting behavior down to lowest $T$. 

In order to interpret this behavior let us recall that for $g_Z<16$ quantum phase slips are no longer bound in pairs. According to the exact solution for the sine-Gordon model  \cite{GNT}, in this case an effective minigap in the spectrum $\tilde \Delta \propto \gamma_{\rm qps}^{\alpha}$ develops implying that at $T \to 0$ samples $Ti3$, $Ti4$ and $Ti5$ should behave as {\it insulators}. In line with these arguments, our resistance data in Fig. 3a demonstrate that the supercurrent in these samples is fully blocked by QPS down to lowest available temperatures and, hence, their insulating behavior should indeed be expected at $T < \tilde \Delta$. The absence of any visible resistance upturn at low $T$ most likely implies that the latter condition is not yet reached and/or the inequality $L \gg L_c$ is not satisfied well enough for these samples. In any event, here superconductivity is totally wiped out by quantum fluctuations in accordance with our theoretical arguments.

Note that the resistance data similar to those of Fig. 3a were also reported previously \cite{BT,Lau,Bezr08} for a large number of $MoGe$  nanowires with shorter values of $\xi$ and $L$. In some of these samples the resistance upturn at lower $T$ indicating the insulating behavior was observed. Reanalyzing the data \cite{BT,Lau,Bezr08} we conclude that they are also consistent with the above physical picture involving the correlation length $L_c$ (\ref{LQPS}), i.e. the superconducting $MoGe$ samples obey the condition $L \lesssim L_c$, whereas the non-superconducting ones typically have the length $L$ exceeding $L_c$. Hence, retrospectively the observations \cite{BT,Lau,Bezr08} also receive a natural explanation which was not yet available at that time.
\begin{figure}
\begin{center}
\includegraphics[width=80mm]{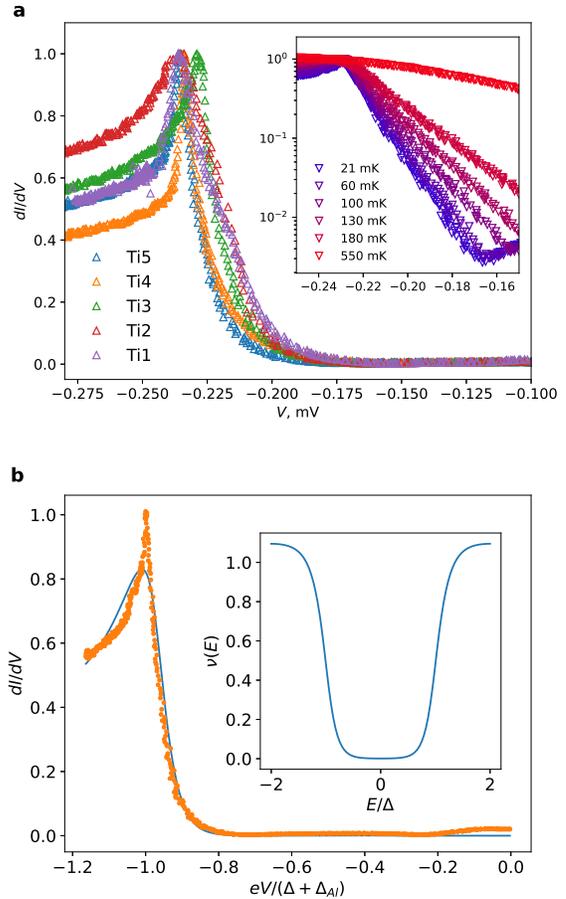}
\end{center}
\caption{{\bf Local differential conductance and electron density of states.} (a) Differential conductance $dI/dV$ as a function of voltage $V$ measured in $Ti-Al$ tunnel junctions at  $T \simeq 21$ mK  for five samples $Ti1$ to $Ti5$. A sharp peak is observed at $e|V|=\Delta + \Delta_{Al}$. Inset: The same data for sample $Ti3$ at different temperatures. 
(b) Fit of the data for sample $Ti3$ at  $T \simeq 21$ mK to the theory \cite{RSZ17}. Inset:
The density of states $\nu$ (in units of the normal density of states at the Fermi energy) as a function of energy $E$ reconstructed for the same sample at the same $T$. Error bars are smaller than data points.}
\end{figure}

{\bf Local properties}

Measurements of the total resistance $R(T)$ alone are not yet sufficient to obtain complete information about the quantum mechanical ground state of superconducting nanowires.  In order to probe their local properties we performed measurements of the $I-V$ curves for tunnel junctions between $Ti$ nanowires and bulk $Al$ electrodes (with the BCS gap $\Delta_{Al}\simeq 190$ $\mu V$), see Fig. 2. The corresponding results for all five samples  are displayed in Fig. 4. In these samples the differential conductance for $Ti-Al$ tunnel junctions has a peak which position varies slightly from sample to sample. As the peak is expected to occur at $e|V|=\Delta + \Delta_{Al}$, we immediately reconstruct the local gap value ranging between $\Delta \approx 50$ $\mu eV$ and $\Delta \approx 37$ $\mu eV$ depending on the sample.  Hence, quantum fluctuations tend to reduce $\Delta$ in superconducting nanowires below its bulk value $\Delta_{Ti} \simeq 60$ $\mu eV$ and this effect appears more pronounced for thinner samples. On the other hand, a non-zero local superconducting gap $\Delta$ remains clearly observable in all our samples. 

As compared to the standard BCS-like $I-V$ curve, systematic broadening of this peak in $dI/dV$ with decreasing wire cross section is observed. This broadening increases with $T$ (cf. inset in Fig. 4a) and it can be explained  \cite{RSZ17,Kostya} if we bear in mind that electrons exchange energies with an effective dissipative environment formed by Mooij-Sch\"on plasmons propagating along the wire. As a result, in our $Ti$ nanowires the singularity in the electron density of states (DOS) $\nu(E)$ at $|E| = \Delta$ and $T \to 0$ gets weaker with decreasing wire cross section and becomes washed out by quantum fluctuations at $g_Z \leq 2$.

This is exactly what we observe in our experiment. By fitting the corresponding $I-V$ data for $Ti-Al$ tunnel junctions to theoretical predictions  \cite{RSZ17} (see Supplementary Note 3) we reconstruct the energy-dependent DOS $\nu(E)$ for our $Ti$ nanowires, as displayed in Fig. 4b.  The best fit for sample $Ti3$ yields the value $g_Z\simeq 1.50$  just slightly below our theoretical estimate  $g_Z\simeq 2.26$. In contrast to the standard BCS dependence $\nu_{\rm BCS}(E) = {\rm Re} |E|/\sqrt{E^2 - \Delta^2}$, here the gap singularities are totally smeared due to electron-plasmon interactions. Nevertheless the superconducting gap in DOS  $\nu(E)$ remains clearly visible. At nonzero $T$ and subgap energies DOS decays exponentially with decreasing $|E|$ as $\nu(E) \propto \exp(-(\Delta-|E|)/T)$ (cf. inset in Fig. 4b) which is also due to interaction between electrons and Mooij-Sch\"on plasmons \cite{RSZ17}.
\begin{figure}
\begin{center}
\includegraphics[width=80mm]{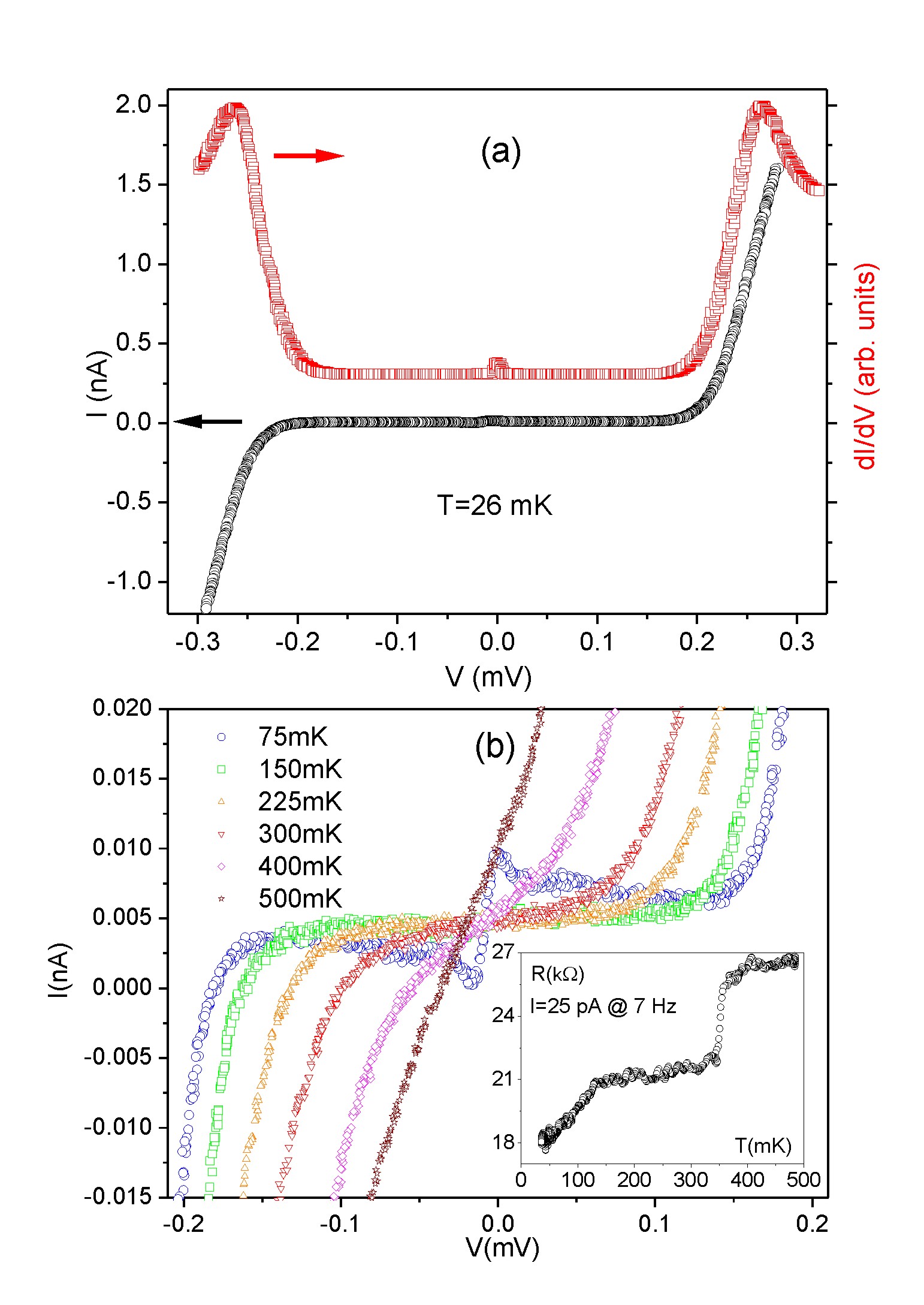}
\end{center}
\caption{{\bf Current-voltage characteristics and Josephson current.} (a) Current-voltage characteristics for $Ti-Al$ tunnel nanojunction corresponding to the $Ti$ nanowire with $L=20$ $\mu$m, $d=35$ nm and $w=38$ nm recorded at $T=26$ mK. (b) Zoom of the current versus voltage dependencies taken at various temperatures. At $T \simeq 75$ mK one clearly observes the Josephson current which gradually disappears at higher $T$. Inset: The total resistance $R$ measured for this nanowire as a function of temperature. Error bars are smaller than data points.}
\end{figure} 

A pronounced superconducting gap in DOS is not the only feature indicating that locally superconducting properties 
remain preserved despite the effect of quantum fluctuations. In Fig. 5 we display the $I-V$ curves measured at different $T$ for yet one more $Ti$ nanowire with local DOS also showing a pronounced superconducting gap (Fig. 5a) and total resistance $R(T)$ behaving qualitatively similarly to that of the samples $Ti3$, $Ti4$ and $Ti5$ (cf. inset  in Fig. 5b). Zooming at the origin of these $I-V$ curves we clearly observe the Josephson current $\sim 5$ pA at $T \simeq 75$ mK, see Fig. 5b. Quite naturally, due to strong fluctuation effects inside our $Ti$ nanowire \cite{RSZ19,RSZ20} this current value is orders of magnitude smaller than the nominal maximum Josephson current of few nA estimated from the standard Ambegaokar-Baratoff formula \cite{Tinkham}. Fluctuation effects become even stronger with increasing temperature and totally wash out the Josephson current already at $T \gtrsim 150$ mK. 

Note that the same Josephson current feature is detected in other $Ti$ nanowires at low $T$ and $V\to 0$, cf., e.g., Fig. 4a (inset) and Fig. 4b. These observations of dc Josephson effect in $Ti-Al$ tunnel junctions further support our conclusion suggesting the presence of local superconductivity in all investigated $Ti$ samples, including the most resistive ones. 

{\bf DISCUSSION}
 
We arrive at the following physical picture describing ultrathin superconducting wires in the "insulating" regime $g_Z <16$ at low enough temperatures. In this regime QPS proliferate while TAPS effects can already be neglected. In thicker nanowires with $L_c \gtrsim L$ (samples $Ti1$ and $Ti2$) quantum phase slips alone cannot disrupt phase coherence across the wire. Such samples then behave to a large extent similarly to effectively zero-dimensional objects, such as, e.g., small-size Josephson junctions with the fluctuating phase \cite{SZ90} embedded in a low resistive external circuit. Depending on the experimental realization \cite{SZ13,RSZ19,RSZ20}, these nanowires may either stay superconducting or become resistive, albeit typically with rather small $R \propto \gamma_{\rm qps}^2$. In contrast, thinner samples with $L_c \ll L$ remain highly resistive with $R \sim R_N$ even at $T \ll T_C$ and should turn insulating in the limit $T \to 0$. This behavior is due to QPS which suppress long range phase coherence in such nanowires.

Remarkably, the superconducting gap $\Delta$ in the energy spectrum of all our $Ti$ nanowires, including highly resistive ones, is reduced but {\it not} destroyed by quantum fluctuations. In addition, this spectrum is also affected by the interaction between electrons and soft phase fluctuation modes (Mooij-Sch\"on plasmons) which washes out the BCS gap singularity in DOS of ultrathin ($g_Z <2$) nanowires and produces a weak subgap tail in $\nu (E)$ at non-zero $T$. We have demonstrated that the wire segments of length $\lesssim L_c$ retain their superconducting properties. On the other hand, longer nanowires composed of many such superconducting segments exhibit effective localization of Cooper pairs at lengths $\sim L_c$ and loose their ability to sustain any measurable supercurrent. These nanowires demonstrate a resistive behavior with $R(T) \sim R_N$ even at $T \ll T_C$ and should turn insulating in the limit of large $L$ and $T \to 0$.

It is well known that under certain conditions granular superconducting arrays and Josephson junction chains may also become resistive and even insulating \cite{PZ89,FS91,BFSZ,FZ}. In that regime superconductivity is well preserved only inside grains while dissipativeless charge transfer across the system is prohibited due to Coulomb blockade of Cooper pair tunneling between such grains. Here, in contrast, we are dealing with nominally uniform nanowires which do not contain any grains and dielectric barriers at all. Nevertheless, such nanowires may exhibit both resistive and insulating behavior as long as their length $L$ strongly exceeds typical size of a "superconducting domain" $L_c \propto \gamma_{\rm qps}^{-\alpha}$. Similarly to normal metallic structures \cite{book,N99,GZ01} this non-trivial feature can be interpreted as weak Coulomb blockade of Cooper pairs that -- as it is illustrated by our results -- may occur even in the absence of tunnel barriers.

In summary, we have demonstrated -- both experimentally and theoretically -- that long and uniform superconducting nanowires in the so-called "insulating" regime actually exhibit a more complicated behavior characterized by superposition of local superconductivity and effective global localization of Cooper pairs. This fundamental property of superconducting nanowires needs to be accounted for while designing various nanodevices with novel functionalities.

{\bf METHODS}

E-beam lift-off process, vacuum deposition of metals and in situ oxidation were used to fabricate tunnel junctions between aluminum electrodes and titanium nanostripes.  Each structure enables one to carry out both pseudo-four-terminal measurements of the total resistance $R(T)$ for all $Ti$ nanowires and local measurements of the $I-V$ curve for all $Al-Ti$ tunnel junctions (Fig. 2). Differential conductances $dI/dV$ were obtained by modulation technique using lock-in amplification. All experiments were made inside $^3He^4He$ dilution refrigerator with carefully filtered \cite{Zav} input/output lines connecting sample to laboratory digital electronics through battery powered analogue pre-amplifiers (see Supplementary Note 4 for details). 

{\bf DATA AVAILABILITY}

The data that support the findings of this study are available from KYA (karutyunov@hse.ru) upon reasonable request.

{\bf ACKNOWLEDGEMENTS}

A.R. acknowledges financial support from RFBR grant No. 19-32-90229. Experimental activity of K.Yu.A. has been supported by the "Mirror Lab" project of HSE University.

{\bf ADDITIONAL INFORMATION}
Supplementary information is available for this paper below.

\newpage


\section*{Supplementary material}

# Superconducting insulators and localization of Cooper pairs

K. Yu. Arutyunov[1,2], J. S. Lehtinen[3], A. A. Radkevich[4], A. G. Semenov[1,4] and A. D. Zaikin[4,5]

[1]*National Research University Higher School of Economics, 101000 Moscow, Russia*
[2]*P.L. Kapitza Institute for Physical Problems RAS, 119334, Moscow, Russia*
[3]*VTT Technical Research Centre of Finland Ltd.,
Centre for Metrology MIKES, P.O. Box 1000, FI-02044 VTT, Espoo, Finland*
[4]*I.E.Tamm Department of Theoretical Physics, P.N.Lebedev Physical Institute, 119991 Moscow, Russia*
[5]*Institute for Quantum Materials and Technologies,
Karlsruhe Institute of Technology (KIT), 76021 Karlsruhe, Germany*

(Dated: April 29, 2021)

## SUPPLEMENTARY INFORMATION

### Supplementary Note 1

In order to fit the resistance data $R(T)$ for samples $Ti1$ and $Ti2$ at temperatures not too far from $T_C$ (albeit outside the critical fluctuation region in its immediate vicinity) we employed the formula for the TAPS contribution to $R(T)$ derived in Ref. [9]. It reads

$$\frac{R_{\text{taps}}}{R_q} = 2A\sqrt{6\pi}\frac{T_C}{T}\frac{L}{\xi(T)}\sqrt{\frac{\delta F}{T}}e^{-\delta F/T}, \quad \text{(S1)}$$

where, as usually, the activation exponent $\delta F/T = \kappa g_\xi (1-t)^{3/2}/t$ with $\kappa \simeq 0.665$ represents an effective free energy barrier for TAPS normalized by temperature, $t = T/T_C$ and $A$ is a dimensionless constant of order one. Rewriting Eq. (S1) in a somewhat more convenient form

$$\frac{R_{\text{taps}}(T)}{R_N} = B\frac{(1-t)^{5/4}}{t^{3/2}}\exp\left(-\kappa g_\xi \frac{(1-t)^{3/2}}{t}\right) \quad \text{(S2)}$$

and fitting the resistance data to the formula

$$R(T) = R_N/(1 + R_N/R_{\text{taps}}(T)), \quad \text{(S3)}$$

which also includes the contribution from overgap quasiparticles, we arrive at excellent fits for both samples $Ti1$ and $Ti2$ displayed in Fig. 3a. The best fit parameters extracted from these fits are $g_\xi \simeq 37.4$, $B \simeq 58.8$ for sample $Ti1$ and $g_\xi \simeq 9.0$, $B \simeq 61.9$ for sample $Ti2$.

### Supplementary Note 2

In order to estimate the correlation length $L_c$ (3) it is necessary to explicitly determine the dimensionless factors $a$ and $b$ entering the QPS amplitude $\gamma_{\text{qps}}$. Note that the effect of TAPS alone is already sufficient to fully explain the dependence $R(T)$ for samples $Ti1$ and $Ti2$ down to low enough temperatures and, hence, QPS effects appear to be of little relevance for these samples. On the other hand, at even lower $T$ the crossover between TAPS- and QPS-dominated regimes should inevitably occur. Judging from our resistance data displayed in Fig. 3a we conclude that for both samples $Ti1$ and $Ti2$ the corresponding crossover temperature $T^*$ can hardly be identified because at such $T$ the resistance values $R(T)$ are already too small to be reliably measured. Still, this observation allows us to estimate the lower bound for $L_c$.

The QPS-controlled contribution to the nanowire resistance $R_{\text{qps}}$ can be described by the following perturbative formula [1,2,10]

$$\frac{R_{\text{qps}}}{R_N} = b^2 g_\xi^3 \frac{v}{16T\xi}\left(\frac{2\pi T}{\Delta}\right)^{g_Z/8-1}\exp(-2ag_\xi). \quad \text{(S4)}$$

The TAPS-QPS crossover temperature $T^*$ is then fixed by the condition

$$R_{\text{qps}} \approx R_{\text{taps}}(T^*). \quad \text{(S5)}$$

From the resistance data for sample $Ti2$ we estimate $R(T^*) \lesssim R_N e^{-4}$ and, hence, $T^* \lesssim T_C e^{-1.15}$. Employing Eqs. (S2), (S4) together with the parameter values $v = 1.91 \cdot 10^6$ m/s, $\xi = 140$ nm, $g_Z = 2.5$ and $g_\xi = 9.0$, we obtain the upper bound for the QPS amplitude (or fugacity). It reads

$$b^2 e^{-2ag_\xi} \lesssim 3.1 \cdot 10^{-7} \quad \text{(S6)}$$

or, equivalently, $ag_\xi - \ln b \gtrsim 7.5$. Making use of Eq. (3) we immediately arrive at the lower bound for the correlation length for this sample $L_c \gtrsim 12\mu m$.

### Supplementary Note 3

Fitting of local differential conductance data $dI/dV$ for $Ti - Al$ tunnel junctions can be performed by means of the standard "semiconductor" formula [6]

$$I(V) \propto \int dE\, \nu(E)\nu_{Al}(E+eV)\Big(F(E) - F(E+eV)\Big), \quad \text{(S7)}$$

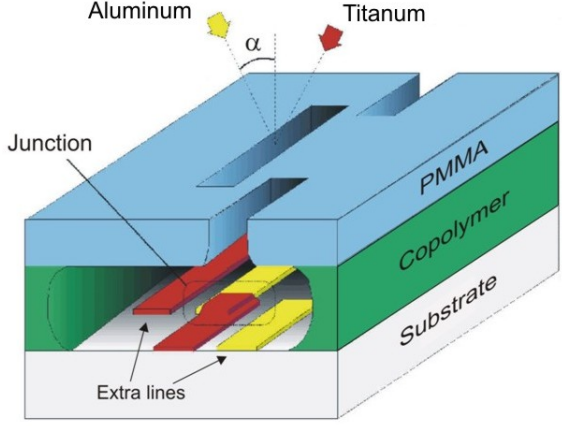

FIG. S1: Schematics of sample fabrication.

where DOS in the bulk aluminum electrode has the usual BCS form $\nu_{Al}(E) = \mathrm{Re}|E|/\sqrt{E^2 - \Delta_{Al}^2}$, whereas the electron DOS in $Ti$ nanowires in the presence of quantum fluctuations reads [34]

$$\nu(E) = \int \frac{d\epsilon}{2\pi} \nu_{BCS}(\epsilon) \mathcal{B}^K(E-\epsilon)\bigl(1+F(\epsilon)F(E-\epsilon)\bigr). \quad (S8)$$

Here $\nu_{BCS}(\epsilon) = \mathrm{Re}|\epsilon|/\sqrt{\epsilon^2 - \Delta^2}$ is the BCS density of states, $F(\epsilon) = \tanh(\epsilon/2T)$ and

$$\mathcal{B}^K(\epsilon) = \cosh\left(\frac{\epsilon}{2T}\right)\left(\frac{2\pi T}{\Delta}\right)^{1/g_Z} \frac{\left|\Gamma\left(\frac{1}{2g_Z} + \frac{i\epsilon}{2\pi T}\right)\right|^2}{2\pi T \Gamma(1/g_Z)}, \quad (S9)$$

where $\Gamma(x)$ is the Euler gamma-function. Combining Eqs. (S7)-(S9) and numerically taking a derivative of $I$ with respect to $V$ we obtain the differential conductance $dI/dV$ employed in order to fit the experimental data for our $Ti$ samples. After proper rescaling these fits are controlled by a single fit parameter $g_Z$ which value can be independently determined by means of this procedure. The corresponding results for sample T3 are displayed in Fig. 4b giving the best fit value $g_Z \simeq 1.50$. By deconvolution of the $I-V$ curve we also determine the electron DOS inside our $Ti$ nanowires, see the inset in Fig. 4b.

## Supplementary Note 4

The nanostructures were fabricated with e-beam lift-off lithography using double layer PMMA/copolymer mask (Fig. S1). Multi-angle e-beam evaporation of metals was made in UHV chamber with residual pressure of gases $< 10^{-8}$ mBar. In-situ oxidation of aluminum in loading chamber at pressure 1 mBar during 1 min formed thin $AlO_x$ layer. Deposition of titanium on top enabled formation of tunnel barrier.

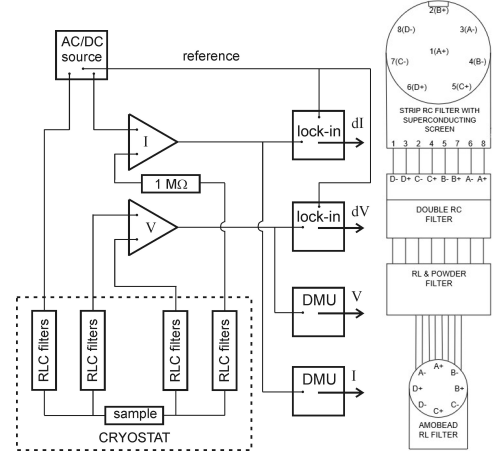

FIG. S2: Left panel: Schematics of measurements. Right panel: Multistage $RLC$ filtering circuit for each sample stage with eight contact lines.

The measurements of electronic transport properties were made at ultra-low temperatures inside $^3He^4He$ refrigerator (Fig. S2). All input/output lines were carefully filtered using multi-stage $RLC$ passive circuit.



\end{document}